\documentclass[12pt]{article}

\ifdefined\pdfminorversion
\fi
\ifdefined\pdfobjcompresslevel
\fi

\usepackage[a4paper,left=2.5cm,right=2.5cm,top=3cm,bottom=3cm]{geometry}
\usepackage{setspace}
\usepackage{amsmath,amssymb,amsfonts,amsthm,mathrsfs}

\DeclareMathOperator*{\argmin}{argmin}

\allowdisplaybreaks[4]

\usepackage{xcolor}
\usepackage{graphicx}
\usepackage[figuresright]{rotating}
\usepackage{booktabs}
\usepackage{multirow}
\graphicspath{{figures/}}

\usepackage{algorithm}
\usepackage[noend]{algpseudocode}

\usepackage{natbib}
\usepackage{xr}
\usepackage{hyperref}
\hypersetup{
    colorlinks=true,
    linkcolor=blue,
    citecolor=blue,
    urlcolor=blue
}

\newtheorem{theorem}{Theorem}

\newtheorem{remark}{Remark}
\newtheorem{definition}{Definition}

\begin{document}

\vspace*{0.5cm}

\begingroup
\singlespacing
\centering

{\fontsize{17}{22}\selectfont\bfseries
Nash Equilibrium Individualized Treatment Regimes
under Network Interference\par}

\vspace{2.0em}

{\normalsize
Junwen Xia$^{1}$ \qquad
Jingxiao Zhang$^{1,*}$ \qquad
Dehan Kong$^{2,*}$\par}

\vspace{0.7em}

{\small
$^{1}$Center for Applied Statistics and School of Statistics,
Renmin University of China, Beijing 100872, China\par
\vspace{0.25em}
$^{2}$Department of Statistical Sciences,
University of Toronto, Toronto, ON M5G 1Z5, Canada\par}

\vspace{0.6em}

{\small
$^{*}$Correspondence:
\href{mailto:zhjxiao@ruc.edu.cn}{zhjxiao@ruc.edu.cn};
\href{mailto:dehan.kong@utoronto.ca}{dehan.kong@utoronto.ca}\par}

\vspace{0.5em}
{\small \today\par}

\par
\endgroup

\vspace{1.2em}

\begingroup
\setstretch{1.2}
\normalsize

\vspace{-0.5em}

\begin{abstract}
    {Network interference creates a fundamental tension in optimal individualized treatment regime learning: maximizing overall population outcomes may come at the expense of some individuals' outcomes. Existing optimal individualized treatment regime methods under interference primarily target population-level objectives without accounting for individual optimality. Motivated by this tension, we introduce Nash equilibrium individualized treatment regimes, or simply Nash regimes, which directly target individual optimality under interference. Under a Nash regime, no individual can improve their conditional expected outcome through unilateral deviation when all other individuals follow the regime. We establish conditions for the existence of Nash regimes. When no exact Nash regime exists, we introduce an \(\alpha^*\)-Nash regime as an alternative target that minimizes regret relative to individual optimality. We then develop Nash learning, a doubly robust procedure for estimating the proposed regimes. Under a super-population random-graph framework accommodating both sparse and dense networks, we characterize how network sparsity and treatment assignment mechanisms affect estimation performance. Simulations and a real-data application demonstrate that Nash learning improves individual optimality while maintaining competitive population-level performance.}
\end{abstract}

\vspace{0.8em}


\par
\endgroup

\vspace{1.2em}

\vfill

\section{Introduction} \label{Sec:introduction}
An optimal individualized treatment regime (OITR) is a decision rule that maps individuals' baseline covariates to treatment assignments with the goal of achieving the best outcomes. Existing approaches generally fall into two broad categories: model-based methods and model-free methods. Model-based methods include Q-learning \cite[]{watkins1992qlearning}, A-learning \cite[]{murphy2003optimal,shi2018highdimensional}, and weighted ordinary least squares (WOLS) \cite[]{wallace2015doubly}. These approaches seek to estimate treatment regimes that optimize individual outcomes by first estimating the conditional average treatment effect (CATE) and subsequently constructing the OITR according to the sign of the CATE.
Model-free methods encompass value search methods \cite[]{zhang2012a} and classification learning \cite[]{zhao2012estimating}. Rather than optimizing individual outcomes, these methods aim to optimize overall population outcomes by first estimating the overall counterfactual mean outcomes under any regime and then employing genetic algorithms or classification techniques to derive the regime that maximizes the overall counterfactual mean outcomes. Despite their different formulations, the two perspectives lead to the same optimal regime under no interference: pointwise optimization of conditional outcomes also maximizes population value.

The no-interference assumption \cite[]{hudgens2008toward} posits that each unit’s outcome is unaffected by the treatment assignments of other units. This assumption is often violated in real-world settings. For example, in vaccine studies, an individual’s outcome may depend not only on their own vaccination status but also on whether those around them are vaccinated \cite[]{halloran1991direct}. When interference exists, improving population-level outcomes may come at the expense of some individuals' outcomes, and vice versa. We refer to scenarios in which such conflicts arise between individual-level and population-level objectives as individual-population disagreement settings. To illustrate this conflict, we present a toy example and relate it to course assignment. An additional example can be found in Section \ref{Sec:addapplication} of the supplementary material.

Consider a fully connected network with outcomes \(Y_i=\bar A_{-i}+A_i\delta(X_i)+\varepsilon_i\), where \(A_i\in\{0,1\}\) is the treatment, \(\bar A_{-i}\) is the fraction of treated individuals excluding unit \(i\), \(X_i\) denotes covariates, and \(\varepsilon_i\) is mean-zero noise. This model consists of positive spillovers \(\bar A_{-i}\) and heterogeneous treatment effects \(\delta(X_i)\). On the one hand, assigning more individuals to treatment \(1\) increases \(\bar A_{-i}\), thereby improving population outcomes through positive spillovers. On the other hand, individuals with \(\delta(X_i)<0\) are harmed by receiving treatment \(1\) and would prefer treatment \(0\). This illustrates that the individual-population disagreement arises from the coexistence of spillovers and heterogeneous treatment effects. Therefore, such a conflict may arise in many applications including course assignment \cite[]{sacerdote2011peer}: the relative benefits of introductory and advanced courses may vary with students' preparedness, while knowledge acquired in advanced courses may generate positive spillovers that benefit peers through informal interactions.
 
When the objective is to maximize population-level outcomes under interference, several approaches are available (e.g., \citealp{su2019modelling,zhang2023individualized,viviano2025policy}).
Among these, \cite{viviano2025policy} estimates overall counterfactual mean outcomes while accounting for network interference and then searches for the regime that maximizes the resulting population value.
Because these approaches target overall population outcomes, the population-optimal regime may come at the expense of some individuals in individual-population disagreement settings. This may make such regimes difficult to implement in practice, as individuals may have incentives to deviate from their assigned treatments, potentially leading to unstable treatment assignments.
Moreover, pursuing maximal overall population outcomes at the expense of some individuals' outcomes is less consistent with the concept of \textit{optimal individualized treatment regime (OITR)} \cite[]{cui2025policy}.

This motivates a complementary objective that directly targets individual-level optimality under interference. Defining such a regime is nontrivial because, under interference, an individual's optimal treatment depends on the treatments assigned to others. A related approach is \cite{jiang2023dynamic}, who define the CATE conditional on individual covariates and the fraction of treated neighbors and derive treatment rules by fixing neighbors' treatments, for example, uniformly at 0 or 1. However, such fixed exposure levels need not coincide with the neighbor treatments induced when a regime is simultaneously implemented across the network. As a result, some individuals may achieve lower outcomes under the assigned treatment than they would by unilaterally deviating, indicating that such treatment assignments do not necessarily optimize individual outcomes.

{In this paper, we propose the first OITR definition designed to optimize individual outcomes under network interference. Specifically, we define individual-level optimality as the condition that no individual can improve their conditional expected outcome by unilaterally deviating from the prescribed treatment, when all other individuals follow the prescribed regime. Because this criterion mirrors the best-response condition underlying Nash equilibrium \cite[]{nash1950equilibrium,fudenberg1991game}, we refer to such regimes as Nash equilibrium individualized treatment regimes, or simply Nash regimes.} By ensuring that no individual can improve their conditional expected outcome through unilateral deviation, a Nash regime may encourage treatment adherence and thereby improve policy effectiveness.
Importantly, although the Nash regime is aimed at achieving individual optimality, our simulation and real-data analyses show that it also achieves high overall population outcomes. 

We summarize the contributions of this paper as follows. 

First, we introduce the Nash equilibrium into the OITR literature and establish conditions for the existence of Nash regimes under network interference. Two interpretable conditions are: (i) treatments are complementary across friends, so that an individual's own treatment effect increases, or at least does not decrease, as more of their friends receive treatment; and (ii) the influence of friends' treatments on an individual's own treatment effect is not dominant. 
Furthermore, when a Nash regime does not exist, we introduce a regret measure that quantifies the deviation of any regime from individual optimality. We then take the \(\alpha^*\)-Nash regime as the target, where \(\alpha^*\) denotes the minimum attainable regret, with \(\alpha^*=0\) whenever a Nash regime exists.

{Second, we develop an OITR framework under a super-population random-graph model. Existing OITR methods under interference typically define optimal regimes relative to particular finite-population networks \citep{su2019modelling,viviano2025policy} or clustered network structures \citep{zhang2023individualized}. Building on the random-graph perspective of \citet{li2022random}, we instead treat the observed network as a realization from an underlying population graph model and target a regime that is transferable across network realizations generated by the same mechanism. To accommodate more realistic and flexible network formation mechanisms, we further extend the random-graph framework of \citet{li2022random} by allowing network formation to depend jointly on observed covariates and latent variables, rather than solely on latent variables. This framework accommodates both sparse and dense networks, thereby relaxing the restriction to sparse networks in existing OITR frameworks.}

Third, we propose a novel method, Nash learning, to estimate the \(\alpha^*\)-Nash regime and establish theoretical guarantees for our method. Since our Nash learning procedure is built upon WOLS \cite[]{wallace2015doubly,jiang2023dynamic}, it is robust to misspecification of either the baseline outcome model or the propensity score model. We also establish how network sparsity levels and treatment assignment mechanisms jointly affect our \(\alpha^*\)-Nash regime estimators. Our analysis reveals that dense networks can be a blessing for estimating the \(\alpha^*\)-Nash regime under a covariate-dependent propensity score, but a curse under a constant propensity score. These are novel theoretical properties for OITR estimation that differ from the existing asymptotic results for causal inference under a super-population random-graph framework \cite[]{li2022random,fan2025causal}. 
En route to establishing these theoretical results, we also derive novel asymptotic properties for WOLS, filling a gap in the theoretical analysis of WOLS in \cite{wallace2015doubly} and \cite{jiang2023dynamic}.

The remainder of this paper is organized as follows. Section \ref{Sec:method} presents the proposed methodology. The theoretical properties of the Nash learning procedure are developed in Section \ref{Sec:theoretical}. Finite-sample performance is examined through simulation studies in Section \ref{Sec:simulation}. Section \ref{Sec:realdata} illustrates the application using social network data. We end with a discussion in Section \ref{Sec:discussion}.

\section{Methodology} \label{Sec:method}
Let \(X\in\mathcal X\) be the \(q\)-dimensional observed covariate vector, \(U\in\mathcal{U}\) be the latent variable vector associated with network generation, \( A \in\{0,1\} \) be a binary treatment, and \( Y \in \mathcal{Y} \) be the outcome. An observed network dataset is denoted as \( D_n=\left(\left\{X_{i}, A_{i}, Y_{i}\right\}_{i=1}^{n}, \boldsymbol E\right) \), where \( \boldsymbol{E} \) is the adjacency matrix of the network and \( n \) is the total number of units. In this paper, we assume \(\{X_i,A_i,U_i\}_{i=1}^n\) are i.i.d. and \(\{Y_i\}_{i=1}^n\) are dependent due to interference. We denote the first-order neighborhood of unit \(i\) by \(\mathcal N_i\), the corresponding treatment vector by \(A_{\mathcal N_i}\), and the covariate vector by \(X_{\mathcal N_i}\). Under network interference, a unit's counterfactual outcome is influenced not only by its treatment but also by its neighbors' treatments, and thus the counterfactual outcome under \((A_i=a_i, A_{\mathcal{N}_i}=a_{\mathcal{N}_i})\) is denoted by \( Y_i\left(a_i, a_{\mathcal{N}_i}\right) \).

We make the following assumptions:

Assumption 1 (Undirected relationships). The network is undirected and has no self-loops: \(E_{ij}=E_{ji}\) for all \(i\neq j\), and \(E_{ii}=0\) for all \(i\).

Assumption 2 (Network generation). There is a symmetric measurable function \( G_n:\mathcal{X}\times\mathcal{U}\times\mathcal{X}\times\mathcal{U}\rightarrow[0,1] \) satisfying \(G_n\left(X_{i},U_i, X_{j},U_j\right)=G_n\left(X_{j},U_j, X_{i},U_i\right)\), such that \( E_{i j} \sim \mathrm{Bernoulli}\{G_n\left(X_{i},U_i, X_{j},U_j\right)\}\) independently for all \( i<j \), where \(U_i\perp\!\!\!\perp A_i|X_i\). 

Assumption 3 (Network sequence). There exists a sequence \(\{\rho_n\}_{n=1}^{\infty}\) such that \(G_n(X_i,U_i,X_j,U_j)\)\\\(=\min\{1,{\rho}_nG(X_i,U_i,X_j,U_j)\}\) for a symmetric measurable function \(G(X_i,U_i,X_j,U_j)\in[\epsilon,1]\) with \(\epsilon>0\) when \(i\neq j\). The sequence \(\{\rho_n\}_{n=1}^{\infty}\) satisfies one of the conditions: \(\rho_n= 1\) for dense network or \(\lim_{n\to\infty}\rho_n=0\) and \(\lim_{n\to\infty}n\rho_n=\infty\) for sparse network.

Assumption 4 (Network consistency).
\(Y_i=Y_i(A_i,A_{\mathcal N_i})\).

Assumption 5 (No unmeasured confounders).
\(Y_i(a_i,a_{\mathcal N_i})\perp\!\!\!\perp (A_i,A_{\mathcal N_i})| (X_i,X_{\mathcal N_i})\).

Assumption 6 (Anonymous interference).
\(Y_i(a_i,a_{\mathcal N_i})=Y_i(a_i,\bar a_{\mathcal N_i})\), where
\(\bar a_{\mathcal N_i}=|\mathcal N_i|^{-1}\sum_{j\in\mathcal N_i}a_j\).

Assumption 7 (Network overlap).
\(P(A_i=a,\bar A_{\mathcal N_i}=\bar a_{\mathcal{N}_i}\mid X_i,X_{\mathcal N_i})>0\)
for all \(i\) and feasible \((a,\bar a_{\mathcal{N}_i})\).

Assumptions 1--3 generalize the random-graph assumptions in the interference literature \cite[]{li2022random,fan2025causal}. Previous works typically assume that the edges \(E_{ij}\) are determined solely by independent latent variables \(U\). We allow the edges \(E_{ij}\) to be determined by both the observed covariates \(X\) and latent variables \(U\), and \(U\) may be dependent on \(X\). This generalization is more suitable for social network data, since connections between two units often depend on observable characteristics, such as homophily in demographics or shared interests, as well as unobserved factors.
Assumptions 4--7 are commonly utilized in the interference literature to ensure identification. More detailed explanations of these assumptions can be found in Section \ref{Sec:assumption17} of the supplementary material.

\subsection{Nash regimes}
We now define Nash regimes in the OITR setting.
{A Nash regime ensures that no individual can improve their conditional expected outcome through unilateral deviation. This individual-level optimality may promote adherence and help policymakers realize the anticipated benefits of an intervention. For instance, our empirical analysis in Section \ref{Sec:realdata} examines how to assign different types of information sessions to farmers in order to enhance their understanding of disaster insurance. Assigning each farmer a session that is optimal may encourage adherence to the assigned intervention and thereby help policymakers strengthen farmers' capacity to cope with potential disasters.}

We first introduce the notation. An individualized treatment regime is \(d:\mathcal{X}\to\{0,1\}\) from the unit covariate space \(\mathcal{X}\) to the treatment space \(\{0,1\}\). 
The counterfactual outcome when the unit \(i\) follows treatment \(A_i=a\) and the neighbors follow the treatment regime \(d\) is \(Y_i(a,\bar{d}_{\mathcal{N}_i})\), where the fraction of treated neighbors
\(\bar{d}_{\mathcal{N}_i}={\sum_{j\in\mathcal{N}_i}d(X_j)}/|\mathcal{N}_i|.\)

For a given regime \(d\), when all other units follow \(d\), the individually optimal treatment for unit \(i\) with covariates \(X_i\) can be defined as the treatment that maximizes the finite-sample conditional expected outcome \(E\{Y_i(a,\bar d_{\mathcal N_i})| X_i\}\) over \(a\in\{0,1\}\). Equivalently, the optimal treatment can be characterized by the sign of the finite-sample conditional treatment effect \(\tau_n(X_i;d)=E\{Y_i(1,\bar d_{\mathcal N_i})-Y_i(0,\bar d_{\mathcal N_i})| X_i\}\), where \(\tau_n(X_i;d)>0\) indicates that treatment 1 is preferred, \(\tau_n(X_i;d)<0\) indicates that treatment 0 is preferred, and \(\tau_n(X_i;d)=0\) indicates indifference between the two treatments. Accordingly, a regime \(d\) achieves individual optimality if every unit is assigned an individually optimal treatment; that is, \(d(X_i)=I\{\tau_n(X_i;d)>0\}\) for all \(X_i\) with \(\tau_n(X_i;d)\neq 0\).

Because the distribution of the fraction of treated neighbors \(\bar d_{\mathcal N_i}\) depends on the sample size \(n\), \(\tau_n(x;d)\) generally varies with \(n\), and hence the individual optimality based on \(\tau_n\) would differ across networks of different sizes. In practice, however, an OITR may be learned from a network of size \(n\) and subsequently applied to a target network of a different size \(m\). This motivates a notion of individual optimality that is not tied to a particular finite network size. We therefore define individual optimality through the conditional treatment effect \(\tau(x;d)=\lim_{n\to\infty}\tau_n(x;d)\), provided the limit exists. This leads to the following definition of a Nash regime.

\begin{definition}
\label{def:nashpopulation}
A regime \(d^*\) is a Nash regime if \(d^*(x)=I\{\tau(x;d^*)>0\}\) for \(P_X\)-almost every \(x\) with \(\tau(x;d^*)\neq 0\).
\end{definition}

Under a Nash regime \(d^*\), the following holds almost surely: no unit \(i\) can improve its conditional expected outcome by unilaterally deviating from the prescribed treatment \(d^*(X_i)\), while all other units continue to follow \(d^*\). Specifically, the change in conditional expected outcome from such a deviation is \(\lim_{n\to\infty}[E\{Y_i(1-d_i^*,\bar d^*_{\mathcal N_i})\mid X_i\}-E\{Y_i(d_i^*,\bar d^*_{\mathcal N_i})\mid X_i\}]=(1-2d_i^*)\tau(X_i;d^*)\leq 0,\) where \(d_i^*=d^*(X_i)\).

\begin{remark}
The Nash regime in Definition \ref{def:nashpopulation} is motivated by the best-response principle underlying Nash equilibrium \citep{nash1950equilibrium,fudenberg1991game}, but represents a different population-level decision object. A standard Nash equilibrium is defined for a given game as a strategy profile \((s_1,\ldots,s_n)\) such that no player benefits from unilateral deviation. In contrast, a Nash regime \(d^*\) is a population-level treatment rule that maps individual covariates to treatment assignments, with individually optimal treatments characterized by conditional expected outcomes.
\end{remark}

\subsection{Existence of Nash regimes} \label{Sec:existence}
In this subsection, we discuss the existence of Nash regimes. Since the Nash regime is defined via the conditional treatment effects \(\tau(x;d)=\lim_{n\to\infty}\tau_n(x;d)\), we first posit some mild conditions to ensure the existence of the limit \(\tau(x;d)\). Note that under the no unmeasured confounders assumption (Assumption 5),
\begin{align} \label{eq:ident_taun}
     \tau_n(x;d)=E\{\delta(X_i,X_{\mathcal{N}_i},\bar{d}_{\mathcal{N}_i})|X_i=x\},
\end{align}
where \(\bar{d}_{\mathcal{N}_i}\) is the fraction of treated neighbors following regime \(d\), and the treatment contrast function \(\delta(X_i,X_{\mathcal{N}_i},\bar{d}_{\mathcal{N}_i})=E(Y_i | X_i, X_{\mathcal{N}_i}, A_i=1, \bar{A}_{\mathcal{N}_i}=\bar{d}_{\mathcal{N}_i})-E(Y_i | X_i, X_{\mathcal{N}_i}, A_i=0, \bar{A}_{\mathcal{N}_i}=\bar{d}_{\mathcal{N}_i})\). 

Because the dimension of \(X_{\mathcal N_i}\) varies across units and with the network size \(n
\), directly analyzing the limit of \(\tau_n(x;d)\) is cumbersome. Following \cite{jiang2023dynamic} and \cite{viviano2025policy}, we assume that \(E(Y_i | X_i, X_{\mathcal{N}_i}, A_i=a, \bar{A}_{\mathcal{N}_i}=\bar{d}_{\mathcal{N}_i})=E(Y_i | X_i, A_i=a, \bar{A}_{\mathcal{N}_i}=\bar{d}_{\mathcal{N}_i})\) such that \(\delta(X_i, X_{\mathcal{N}_i}, \bar{d}_{\mathcal{N}_i}) = \delta(X_i, \bar{d}_{\mathcal{N}_i})\). More generally, one could allow $\delta(X_i, X_{\mathcal{N}_i}, \bar{d}_{\mathcal{N}_i}) = \delta\{X_i, m(X_{\mathcal{N}_i}), \bar{d}_{\mathcal{N}_i}\}$ for some aggregation function $m$ \cite[]{shi2023multiagent} to allow a more flexible specification of the treatment contrast function \(\delta\). We provide a discussion in Section \ref{Sec:proofexistence} of the supplementary material.

The following theorem establishes the existence of \(\tau(x;d)=\lim_{n\to\infty}\tau_n(x;d)\) under boundedness and continuity of the treatment contrast function.

\begin{theorem}\label{the:existencetau}
Under Assumptions 1-7, assume that \(\delta(X_i,\bar{d}_{\mathcal{N}_i})\) is a bounded and continuous function of \(\bar{d}_{\mathcal{N}_i}\) for any \(X_i\in\mathcal{X}\). Then, the limit \(\tau\) exists and 
\[\tau(x;d)=\lim_{n\to\infty} \tau_n(x;d)=E\left[\delta\{X_i, \bar{d}(X_i,U_i)\}|X_i=x\right],\] 
where 
\(\bar d(X_i,U_i)\) denotes the probability limit of the fraction of treated neighbors \(\bar d_{\mathcal{N}_i}\), conditional on \((X_i,U_i)\).
The explicit form of \(\bar{d}\) is provided in (\ref{eq:bard}) of the supplementary material.
\end{theorem}

Next, we discuss the existence of Nash regimes. Note that the Nash regime \(d^*(x)=I\{\tau(x;d^*)>0\}\) corresponds to a fixed point of the mapping
$d \mapsto I\{\tau(\cdot;d)>0\}$. Therefore, we provide two interpretable sufficient conditions for the existence of Nash regimes using the Knaster-Tarski fixed point theorem and the Banach fixed point theorem, respectively. 

\begin{theorem} \label{the:nashexistence}
    Under the assumptions in Theorem \ref{the:existencetau}, Nash regimes exist when one of the following conditions holds.

    (i) The treatment contrast function \(\delta(X_i,\bar{d}_{\mathcal{N}_i})\) is a non-decreasing function of \(\bar{d}_{\mathcal{N}_i}\) for any \(X_i\in\mathcal{X}\).

    (ii) There exists a constant \(\lambda<1\) such that for any regimes \(d_1,d_2\), \[E|I\{\tau(X;d_1)>0\}-I\{\tau(X;d_2)>0\}|\leq \lambda E|d_1(X)-d_2(X)|.\]
    \end{theorem}

Condition (i) requires that the treatment contrast function \(\delta(X_i,\bar{d}_{\mathcal{N}_i})\) does not decrease as the fraction of treated neighbors \(\bar{d}_{\mathcal{N}_i}\) increases. Since \(\delta\) represents the treatment contrast, condition (i) indicates that a Nash regime exists when complementarities arise among friends' treatments, i.e., an individual's own treatment effect increases, or at least does not decrease, as more of their friends receive treatment. 

Condition (ii) requires that small changes in the regime \(d\) do not substantially alter the conditional treatment effect \(\tau(X;d)\). This condition is plausible when changes in the fraction of treated neighbors, \(\bar d_{\mathcal N_i}\), have only a limited effect on the individual's treatment contrast, \(\delta(X_i,\bar d_{\mathcal N_i})\). An extreme case is when the treatment contrast is completely unaffected by the fraction of treated neighbors, so that \(\delta(X_i,\bar d_{\mathcal N_i})=\delta(X_i)\). In this setting, condition (ii) holds with \(\lambda=0\), thereby guaranteeing the existence of a Nash regime.

We further impose an additive model on \(\delta(X_i,\bar d_{\mathcal N_i})\) to gain further insight into the existence of Nash regimes; see Section \ref{Sec:modelassump} of the supplementary material.

\begin{remark}
{Theorem \ref{the:nashexistence} does not guarantee the uniqueness of the Nash regime. When multiple Nash regimes exist, we regard any of them as a valid target. This choice is analogous to the standard mean OITR without interference \citep{luedtke2016statistical}, where the mean OITR may not be unique. One could further select among multiple Nash regimes according to their overall population values, which we leave for future research.}
\end{remark}

\subsection{Generalization to \(\alpha^*\)-Nash regimes}
{While Theorem \ref{the:nashexistence} provides sufficient conditions for the existence of a Nash regime, such a regime is not guaranteed to exist in all settings. We therefore introduce the \(\alpha^*\)-Nash regime as an alternative target when an exact Nash regime does not exist.}


For a regime \(d\), define the finite-sample conditional regret for a unit \(i\) with covariates \(X_i=x\) as the loss from following \(d(x)\) rather than the individual optimal treatment. Specifically, \(r_n(x;d)=\max_{a\in\{0,1\}}E\{Y_i(a,\bar d_{\mathcal{N}_i})| X_i=x\}-E[Y_i\{d(X_i),\bar d_{\mathcal{N}_i}\}| X_i=x]\).
Using the definition of the finite-sample conditional treatment effect \(\tau_n(x;d)\), this regret can equivalently be written as \(r_n(x;d)=-\min\{d(x)\tau_n(x;{d}),0\}-\min\{(d(x)-1)\tau_n(x;{d}),0\}\).
We then define the conditional regret as \(r(x;d)=\lim_{n\to\infty}r_n(x;d)\), and the regret as \(\mathcal R(d)=E\{r(X;d)\}\). 

From Definition \ref{def:nashpopulation}, a regime \(d^*\) is a Nash regime if and only if \(r(X;d^*)=0\), \(P_X\)-almost surely. Since \(r(x;d)\geq0\) for any regime \(d\), this is equivalent to \(\mathcal R(d^*)=E\{r(X;d^*)\}=0\). Thus, \(\mathcal R(d)\) quantifies the extent to which regime \(d\) departs from individual optimality, with smaller values indicating less deviation from individual optimality. This characterization motivates the following definition of an \(\alpha^*\)-Nash regime.

\begin{definition} \label{def:epsilonnash}
    Denote the minimum attainable regret as \(\alpha^*=\min_d\mathcal{R}(d)\). A regime \(d^*\) is an \(\alpha^*\)-Nash regime if the regret \(\mathcal{R}(d^*)=\alpha^*\). When \(\alpha^*=0\), \(d^*\) is a Nash regime.
\end{definition}

{When \(\alpha^*>0\), the positive regret represents the irreducible deviation from exact individual optimality because no regime can simultaneously satisfy the individual optimal treatment condition for the entire population. Since \(r(X;d^*)\geq 0\), Markov's inequality gives \(P\left\{r(X;d^*)\leq {\alpha^*}/{\delta}\right\}\geq 1-\delta\) for any \(\delta\in(0,1)\). Thus, a small \(\alpha^*\) implies that, although exact individual optimality is unattainable, most individuals incur only a small loss relative to their individual optimal treatments. 
}

\subsection{Nash learning for estimating the \(\alpha^*\)-Nash regime} \label{Sec:nashlearning}
In this subsection, we propose Nash learning for estimating the \(\alpha^*\)-Nash regime. The procedure relies on estimators of the conditional treatment effect \(\hat{\tau}(x;d)\) and regret \(\widehat{\mathcal{R}}(d)\), whose construction is presented in Section \ref{Sec:nuisance}. For now, we take these estimators as given and focus on the optimization procedure for learning the regime.


{Because an \(\alpha^*\)-Nash regime minimizes \(\mathcal{R}(d)\), a direct approach to estimating such a regime is to replace \(\mathcal{R}(d)\) with its estimator \(\widehat{\mathcal{R}}(d)\) and solve the empirical regret minimization problem \(\arg\min_{d\in\mathcal{D}_L}\widehat{\mathcal{R}}(d)\), where \(\mathcal{D}_L\) is a prespecified set of candidate regimes. Although the \(\alpha^*\)-Nash regime is defined over the unrestricted regime space, estimation is performed over \(\mathcal D_L\) for computational tractability. 
In this paper, we choose the sieve-induced regime class \(\mathcal{D}_L=\{I\{\Phi(x)^\top \eta>0\}:\|\eta\|_2=1\}\) to accommodate nonlinear regimes, where \(\Phi(x)=(\phi_{1}(x),...,\phi_{L}(x))^\top\) is a vector consisting of \(L\) sieve basis functions, such as splines for continuous covariates or indicator bases for discrete covariates \cite[]{shi2022statistical}. Restricting the optimization to \(\mathcal{D}_L\) introduces a sieve approximation error, which would vanish as \(L\to\infty\).

Since the optimization problem \(\arg\min_{d\in\mathcal{D}_L}\widehat{\mathcal{R}}(d)\) is nonsmooth with respect to \(\eta\), we employ evolution strategies (ES) to solve it. ES is a derivative-free evolutionary optimization method that is well suited to nonsmooth objective functions and scales effectively to complex regime classes involving a large number of coefficients \cite[]{salimans2017evolution}. Further details of the ES algorithm are provided in Section \ref{Sec:ES} of the supplementary material.
Because minimizing the empirical regret over \(\mathcal D_L\) leads to a nonconvex optimization problem and ES can be computationally intensive, we use best-response dynamics, a classical procedure widely used in game theory \citep{gilboa1991social}, to obtain a low-regret regime as a warm start for ES. 
Nash learning therefore consists of two stages. The first stage employs best-response dynamics to obtain a low-regret candidate regime, and the second stage uses this regime as a warm start for ES to directly minimize \(\widehat{\mathcal R}(d)\) over \(\mathcal D_L\).

To motivate the first stage, consider the case in which a Nash regime exists. By Definition \ref{def:nashpopulation}, the Nash regime \(d^*(x)=I\{\tau(x;d^*)>0\}\) is a fixed point of the mapping \(d\mapsto I\{\tau(\cdot;d)>0\}\). This motivates the best-response dynamics, which iteratively updates the regime by choosing the estimated best response to the current regime \(d_t\), namely, \(d_{t+1}(x)=I\{\hat{\tau}(x;d_t)>0\}\). Here, the estimator of the conditional treatment effect, \(\hat{\tau}(x;d_t)\), is constructed using the sieve basis \(\Phi(x)\), so that \(d_{t+1}\in\mathcal D_L\).
Although this iteration is motivated by the fixed-point characterization of an exact Nash regime, it can also provide a useful low-regret candidate when an exact Nash regime does not exist, as observed in our numerical studies. We terminate the iteration at \(d_T\) if either \(\widehat{\mathcal{R}}(d_T)\leq\kappa\) or \(\widehat{\mathcal{R}}(d_{T+1})>\widehat{\mathcal{R}}(d_T)\), and use \(d_T\) as the warm start for the subsequent ES optimization. We recommend setting \(\kappa\) to \(0.001\) times the sample standard deviation of the outcome \(Y\), which performs well in our numerical studies.
The complete Nash learning procedure is summarized in Algorithm \ref{alg:2}. 
}

\begin{algorithm}[!htbp]
\caption{Nash learning for estimating the \(\alpha^*\)-Nash regime.} \label{alg:2} 
\begin{algorithmic}[1]
\State Input the observed dataset \((\{Y_i,A_i,X_i\}_{i=1}^n,\boldsymbol{E})\).
\State Start from an initial regime \(d_{0}\).
\State At iteration \(t\), estimate the conditional treatment effect \(\tau(x;d_t)\) using \(\hat{\tau}(x;d_t)\) in \eqref{eq:estimatortau}, and estimate the regret \(\mathcal{R}(d_t)\) using \(\widehat{\mathcal{R}}(d_t)\) in \eqref{eq:est_regret}.
\State Update:
\begin{align*}
    &d_{t+1}(x) = I\{ \hat{\tau}(x;{d}_{t}) > 0 \}.
\end{align*}
\State Repeat until $ \widehat{\mathcal{R}}(d_T)\leq \kappa$, or \(\widehat{\mathcal{R}}(d_{T+1})>\widehat{\mathcal{R}}(d_T)\).
\State Run the ES algorithm in Section \ref{Sec:ES} of the supplementary material using \(d_T\) as the warm start to obtain  $\hat{d}=\argmin_{d\in\mathcal{D}_L}\widehat{\mathcal{R}}(d)$.
\end{algorithmic}
\end{algorithm}


\begin{remark}
Here, \(\widehat{\mathcal R}(\hat d)\) can be used to quantify the empirical departure of the estimated regime \(\hat d\) from individual optimality.
\end{remark}

\subsection{Implementation details} \label{Sec:nuisance}
Algorithm \ref{alg:2} requires estimators of the conditional treatment effect \(\tau(x;d)\) and the regret \(\mathcal R(d)\). In this subsection, we describe how these estimators are constructed.

Given an estimator \(\hat\tau(x;d)\) of \(\tau(x;d)\), the regret \(\mathcal{R}(d)=-E[\min\{d(X)\tau(X;{d}),0\}+\min\{(d(X)-1)\tau(X;{d}),0\}]\) can be estimated by the plug-in estimator:
\begin{align} 
\label{eq:est_regret}
    \widehat{\mathcal{R}}(d)=-\frac{1}{n}\sum_{i=1}^n\left[\min\{d(X_i)\hat{\tau}(X_i;d),0\}+\min\{(d(X_i)-1)\hat{\tau}(X_i;d),0\}\right].
\end{align}

Thus, it remains to provide an estimator of \(\tau\). Recall that the conditional treatment effect \(\tau(x;d)=\lim_{n\to\infty} \tau_n(x;d)\). Therefore, we propose to estimate \(\tau_n(x;d)\) to approximate \(\tau(x;d)\). The finite-sample conditional treatment effect \(\tau_n\) can be identified by 
\(
    \tau_n(x;{d})=E\{\delta(X_i,\bar{d}_{\mathcal{N}_i})|X_i=x\},
\)
where \(\delta(X_i,\bar{d}_{\mathcal{N}_i})=E(Y_i | X_i, A_i=1, \bar{A}_{\mathcal{N}_i}=\bar{d}_{\mathcal{N}_i})-E(Y_i | X_i, A_i=0, \bar{A}_{\mathcal{N}_i}=\bar{d}_{\mathcal{N}_i})\) is the treatment contrast function. 
Therefore, we can estimate \(\tau_n(x;d)\) in two steps. First, we estimate the treatment contrast function \(\delta\) by treating \(Y_i\) as the outcome and \((X_i,A_i,\bar{A}_{\mathcal{N}_i})\) as covariates, yielding the estimator \(\hat{\delta}\). Then, we use \(\hat{\delta}(X_i,\bar{d}_{\mathcal{N}_i})\) as the outcome to estimate \(\tau_n(x;d)\).


In particular, we first estimate the treatment contrast function 
\(\delta\) using WOLS \cite[]{wallace2015doubly,jiang2023dynamic}. This approach has a doubly robust property: when the model for the treatment contrast function \(\delta\) is correct, the estimator \(\hat{\delta}\) remains consistent if either the baseline outcome model for \(\mu(x,\bar{a})=E(Y_i|X_i=x,A_i=0,\bar{A}_{\mathcal{N}_i}=\bar{a})\) or the propensity score model for \(\pi(a|x)=P(A_i=a|X_i=x)\) is correctly specified. 

Specifically, we assume that the treatment contrast function admits
\(
{\delta}(X_i,\bar{A}_{\mathcal{N}_i})=\Psi(X_i,\bar{A}_{\mathcal{N}_i})^\top{\beta}^*\), 
where
\(
\Psi(X_i,\bar{A}_{\mathcal{N}_i})=(1,X_i^\top,\bar{A}_{\mathcal{N}_i},X_i^\top\bar{A}_{\mathcal{N}_i})^\top
\). This parametric specification is widely adopted in the interference literature for modeling heterogeneous treatment effects and, compared with nonparametric alternatives, provides more stable extrapolation of \(\delta(X_i,\bar d_{\mathcal N_i})\) across different regimes \(d\) \cite[]{jiang2023dynamic,viviano2025policy}. We then estimate \(\beta^*\) by
\begin{align} \label{eq:estimatorWOLS}
 (\hat{\beta},\hat{\gamma})=\argmin_{\beta,\gamma} \sum_{i=1}^{n} \hat{w}(A_i|X_i)\{Y_i-\Psi (X_i,\bar{A}_{\mathcal{N}_i})^\top\gamma-A_i\Psi(X_i,\bar{A}_{\mathcal{N}_i})^\top\beta\}^2.
\end{align}
Here, \(\Psi (X_i,\bar{A}_{\mathcal{N}_i})^\top\gamma\) is used to model the baseline untreated outcome function \(\mu(X_i,\bar{A}_{\mathcal{N}_i})\), and $\hat{w}(A_i | X_i)$ denotes an estimator of the weighting function $w(A_i | X_i)$, satisfying the balancing condition $w(1 | X)\pi(1 | X)=w(0 | X)\pi(0 | X)$. In particular, we adopt the choice $w(A_i | X_i)=|A_i-\pi(1 | X_i)|$, with corresponding estimator $\hat{w}(A_i | X_i)=|A_i-\hat{\pi}(1 | X_i)|$, where $\hat{\pi}(1 | X_i)$ denotes an estimator of the propensity score. The propensity score estimator can be obtained through logistic regression. 



Next, we discuss the estimation of the finite-sample conditional treatment effect \(\tau_n(x;d)=E\{\delta(X_i,\bar{d}_{\mathcal{N}_i})|X_i=x\}\). Our strategy is to first construct a pseudo outcome \(\widehat{Y}_i(d)=\hat{\delta}(X_i,\bar{d}_{\mathcal{N}_i})=\Psi(X_i,\bar{d}_{\mathcal{N}_i})^\top\hat{\beta}\), and then regress this pseudo-outcome on \(X_i\) using a sieve basis. 
{We use sieve regression for two reasons. 
First, \(\bar d_{\mathcal{N}_i}\) converges in probability to \(\bar d(X_i,U_i)\) as \(n\to\infty\), and the resulting regression function \(E[\delta\{X_i,\bar d(X_i,U_i)\}|X_i]\) can be highly nonlinear in \(X_i\), because the limiting neighborhood exposure \(\bar d(X_i,U_i)\) may itself depend on \(X_i\) in a complex way.
More details can be found in Theorem \ref{the:existencetau}. Second, estimating \(\tau\) in the sieve space ensures that the induced regime \(I\{\hat\tau(x;d)>0\}\) belongs to \(\mathcal D_L\), which is essential for implementing the warm start procedure described in Section \ref{Sec:nashlearning}. } 

Specifically, we approximate \(\tau_n(X;d)\) using a linear sieve, \(\tau_n(X;d)\approx\Phi(X)^\top\eta^*(d)\),
where \(\Phi(X)=(\phi_{1}(X),...,\phi_{L}(X))^\top\) is a vector of basis functions discussed in Section \ref{Sec:nashlearning}.
The parameters \(\eta^*(d)\) can be estimated by the least squares method with \(\widehat{Y}_i(d)\) as the outcome. Denote the matrix \(\boldsymbol{\Phi}=(\Phi({X}_1),...,\Phi({X}_n))^\top\) and the Gram matrix \(\boldsymbol{G}_{\Phi}=\boldsymbol{\Phi}^\top\boldsymbol{\Phi}\). The estimate of \({\eta}^*(d)\) then is \(\hat{\eta}(d)={\boldsymbol{G}}_{\Phi}^{-1}\boldsymbol{\Phi}^\top\widehat{Y}(d)\). 
The estimator of \(\tau_n\), and hence of our target \(\tau=\lim_{n\to\infty}\tau_n\), is constructed as follows: 
\begin{align}\label{eq:estimatortau}
    \hat{\tau}(X;{d})=\Phi(X)^\top\hat{\eta}(d).
\end{align}

\section{Theoretical results} \label{Sec:theoretical}

In this section, we establish theoretical guarantees for Nash learning. We first derive the convergence rate of the WOLS estimator \(\hat\beta\), and then propagate this estimation error to the excess regret \(\mathcal R(\hat d)-\min_d \mathcal R(d)\).

{Note that the behavior of the WOLS estimator depends on whether the regressors used to model the treatment contrast remain sufficiently distinct as the network grows. Recall that the regressor vector used in WOLS is \(\Psi(X_i,\bar A_{\mathcal N_i})=(1,X_i^\top,\bar A_{\mathcal N_i},X_i^\top\bar A_{\mathcal N_i})^\top\). Under a constant propensity score, these regressors may become nearly collinear in large networks. For example, when \(\pi(A_i=1| X_i)=0.5\), Lemma \ref{lemma:barpibard} of the supplementary material shows that \(\bar A_{\mathcal N_i}\to_p 0.5\) for every unit \(i\). Consequently, the spillover-related regressors \((\bar A_{\mathcal N_i},X_i^\top\bar A_{\mathcal N_i})^\top\) become approximately \(0.5(1, X_i^\top)^\top\), implying near-linear dependence between the spillover-related regressors and the baseline regressors \((1, X_i^\top)^\top\). This leads to an asymptotic collinearity problem in the WOLS design. For finite \(n\), the design is generally only nearly collinear because \(\bar A_{N_i}\) retains some cross-sectional variation, although this variation vanishes and exact collinearity emerges in the limit. Interestingly, as shown in Section \ref{Sec:existencemulti}, network sparsity can alleviate this near-collinearity problem by retaining greater variation in \(\bar A_{\mathcal N_i}\).}


\subsection{Well-conditioned WOLS design} \label{Sec:absence}

We first consider settings in which the WOLS design remains well-conditioned as the network size increases. By Lemma \ref{lemma:barpibard} of the supplementary material, \(\bar A_{\mathcal N_i}\to_p \bar\pi(X_i,U_i),\) where \(\bar\pi(X_i,U_i)\) denotes the limiting fraction of treated neighbors for a unit with characteristics \((X_i,U_i)\). Consequently, for large networks, the WOLS regressor vector \(\left(1,X_i^\top,\bar A_{\mathcal N_i},X_i^\top\bar A_{\mathcal N_i}\right)^\top\) is approximately \((1,X_i^\top,\bar\pi(X_i,U_i),X_i^\top\bar\pi(X_i,U_i))^\top.\) Accordingly, we impose the following eigenvalue condition to ensure that the WOLS design remains well-conditioned.

Assumption 8: Let \(\Psi\{X_i,\bar\pi(X_i,U_i)\}=(1,X_i^\top,\bar\pi(X_i,U_i),X_i^\top\bar\pi(X_i,U_i))^\top\). There exists a constant \(c_1>0\) such that the minimum eigenvalue satisfies \(\lambda_{\min}(E[\Psi\{X_i,\bar\pi(X_i,U_i)\}\)\\\(\Psi\{X_i,\bar\pi(X_i,U_i)\}^\top])\geq c_1\).

Assumption 8 is plausible for covariate-dependent propensity scores because \(\bar\pi(X_i,U_i)\) varies with \(X_i,U_i\) without imposing special structures on $G(X_i,U_i,X_j,U_j)\pi(A_j=1|X_j)$.

To establish the convergence results, we further impose standard margin, smoothness, positivity, design, and regularity conditions. In particular, the \(\alpha^*\)-Nash regime decision boundary satisfies a margin condition with exponent $\nu$, the relevant regression functions possess H\"older smoothness $p$, and the propensity-score estimator converges at rate
$r_\pi$ to a fixed function $\pi^0$, which may differ from the propensity score $\pi$. We additionally impose regularity conditions on the covariate distribution, baseline outcome function, residuals, and empirical-regret optimization. These conditions are summarized as Assumptions 9--13 in Section \ref{Sec:assumption913} of the supplementary material. With these assumptions, we can derive the convergence rates of \(\|\hat{\beta}-\beta^*\|_2\) and \(\mathcal{R}(\hat{d})-\min_d\mathcal{R}(d)\) as follows:
\begin{theorem} \label{the:asymptotic1}
  Under Assumptions 1--13, suppose that \(L/\log n\to\infty\), \(L\sqrt{\log n}/\sqrt{n\rho_n}\to 0\), and \(n\rho_n/\log n \to\infty\) as \(n\to\infty\). If either the model for the propensity score \(\pi(A|X)\) or the model for the baseline outcome function \(\mu(X,\bar{A}_{\mathcal{N}})\) is correctly specified, then 
  
  (i) \(\|\hat{\beta}-\beta^*\|_2=O_p(n^{-1/2}+r_{\pi})\);
  
  (ii) \(\mathcal{R}(\hat{d})-\min_d\mathcal{R}(d)=O_p(L^{-\min\{\nu,1\} p/q}+L\sqrt{\log n}/\sqrt{n\rho_n}+r_{\pi}).\)
\end{theorem}

Theorem \ref{the:asymptotic1} shows that the effective sample size is \(n\) for estimating the treatment contrast parameter \(\beta^*\), whereas it is \(n\rho_n\) for estimating the \(\alpha^*\)-Nash regime. The latter arises because regime estimation relies on the convergence of the finite-neighborhood exposure \(\bar d_{\mathcal N_i}\) to its population limit \(\bar d(X_i,U_i)\), whose accuracy is governed by the number of neighbors \(|\mathcal N_i|=O_p(n\rho_n)\). Consequently, under covariate-dependent propensity scores, denser networks can improve the estimation of Nash regimes by increasing the effective neighborhood size. This benefit does not generally extend to constant propensity scores, under which dense networks may induce near-collinearity, as discussed in Section \ref{Sec:existencemulti}. Simulation results illustrating this difference are provided in Section \ref{Sec:diffsparsity}.

Theorem \ref{the:asymptotic1} also establishes the doubly robust property of our method. Provided that either the propensity score model \(\pi\) or the baseline outcome model \(\mu\) is correctly specified, \(\hat\beta\) is a consistent estimator of \(\beta^*\) and the regret \(\mathcal R(\hat d)\) converges to the optimal regret \(\min_d\mathcal R(d)\). In particular, when a Nash regime exists so that \(\min_d\mathcal R(d)=0\), the regret of the estimated regime converges to zero in probability.

The excess regret rate consists of three components. The term \(L^{-\min\{\nu,1\}p/q}\) is the sieve approximation error, \(L\sqrt{\log n}/\sqrt{n\rho_n}\) is the stochastic error determined by the sieve dimension and the effective neighborhood size, and \(r_\pi\) reflects the estimation error of the propensity score. The optimal convergence rate for the excess regret \(\mathcal{R}(\hat{d})-\min_d\mathcal{R}(d)\) is $O_p\{(\frac{\log n}{n\rho_n})^{{\min\{\nu,1\}p}/[{2(q+\min\{\nu,1\}p)]}}+r_{\pi}\}$ given that $L\asymp(\frac{n\rho_n}{\log n})^{{1}/[{2(1+\min\{\nu,1\}p/q)}]}.$

\subsection{Near-collinearity under a constant propensity score} \label{Sec:existencemulti}
{In this subsection, we consider the constant propensity score setting, where the WOLS design suffers from the near-collinearity described in Section \ref{Sec:theoretical}. We retain the full set of regressors \((1, X_i^\top,\bar A_{\mathcal N_i}, X_i^\top\bar A_{\mathcal N_i})^\top\) in WOLS and instead investigate how network sparsity alleviates the near-collinearity issue and affects the convergence rates of our estimators. To isolate this near-collinearity issue, we impose Assumption 14 to ensure that the baseline regressors \((1,X_i^\top)^\top\) are themselves well-conditioned.}

Assumption 14: The minimum eigenvalue \(\lambda_{\min}(E\{(1,X_i^\top)^\top(1,X_i^\top)\})\geq c_1\) for some constant \(c_1>0\).

With Assumption 14, we can establish the convergence rate of \(\|\hat{\beta}-\beta^*\|_2\) and \(\mathcal{R}(\hat{d})-\min_d\mathcal{R}(d)\) under constant propensity scores as follows:
\begin{theorem} \label{the:asymptotic2}
  Under Assumptions 1--7, 9--10, 12--14, suppose that, as \(n\to\infty\), \(L/\log n\to\infty\), \(L\sqrt{\log n}/\sqrt{n\rho_n}\to 0\), \(n\rho_n/\log n\to \infty\), and the network is sparse such that \(\rho_n\to0\). If the propensity score is known as a constant, then

(i): \(\|\hat{\beta}-\beta^*\|_2=O_p(\sqrt{\rho_n})\);
  
 (ii) \(\mathcal{R}(\hat{d})-\min_d\mathcal{R}(d)=O_p(L^{-\min\{\nu,1\} p/q}+L\sqrt{\log n}/\sqrt{n\rho_n}+\sqrt{\rho_n}).\)
\end{theorem}

Compared to Theorem \ref{the:asymptotic1}, where the propensity score $\pi(A | X)$ varies with $X$, the convergence rates of $\|\hat{\beta}-\beta^*\|_2$ and $\mathcal{R}(\hat{d}) - \min_d \mathcal{R}(d)$ in Theorem \ref{the:asymptotic2} include an additional term $\sqrt{\rho_n}$ under a constant propensity score. This additional term arises from the near-collinearity induced by the constant propensity score. The term \(\sqrt \rho_n\) in Theorem \ref{the:asymptotic2} shows that, within the sparse-network setting, increasing network density aggravates the near-collinearity issue. This suggests that the near-collinearity issue can be particularly severe in dense networks, a phenomenon we further illustrate in Section \ref{Sec:diffsparsity}. 
The optimal convergence rate for the excess regret \(\mathcal{R}(\hat{d})-\min_d\mathcal{R}(d)\) is 
$
O_p\{(\frac{\log n}{n\rho_n})^{
{\min\{\nu,1\}p}/[
{2(q+\min\{\nu,1\}p)]}
}
+\sqrt{\rho_n}\}$
 given that $L\asymp (\frac{n\rho_n}{\log n})^{
{1}/[{2(1+\min\{\nu,1\}p/q)}]
}.$ 
We also provide simulation results in Section \ref{Sec:diffsparsity} to support Theorem \ref{the:asymptotic2}.



\section{Simulation} \label{Sec:simulation}
\subsection{Results under different propensity scores} \label{Sec:simconstantps}
In this subsection, we investigate the performance of our proposed method under both covariate-dependent and constant propensity scores through simulation studies. Specifically, the covariates $X_i=(X_{i,1}, X_{i,2})$ are two-dimensional, each drawn independently from a standard normal distribution. The covariate-dependent propensity score is set as \(\pi(A_i=1|X_i)=0.8I(X_{i,1}>0)+0.2I(X_{i,1}\leq 0)\), which corresponds to a stratified experimental design, while the constant propensity score is set to the known constant \(\pi(A_i=1|X_i)=0.5\). The network is generated by the function \(G_n(X_i,U_i,X_j,U_j)=\rho_nG(X_i,U_i,X_j,U_j)\) with \(\rho_n=n^{-1/3}\) corresponding to a sparse network and several specifications of \(G\) as follows:

(i): \(G(X_i,U_i,X_j,U_j)=\mathrm{expit}(-1-|X_{i,1}-X_{j,1}|)\);

(ii): \(G(X_i,U_i,X_j,U_j)= 0.3I\{\mathrm{sign}(X_{i,1})=\mathrm{sign}(X_{j,1})\}  +0.02I\{\mathrm{sign}(X_{i,1})\neq \mathrm{sign}(X_{j,1})\}\);

(iii): \(G(X_i,U_i,X_j,U_j)=\mathrm{expit}(-1-|X_{i,1}-X_{j,1}|-|U_{i}-U_{j}|)\), where \(U_i\) is an unobserved variable independently generated from uniform distribution on \([0,1]\).

Network (i) is an observed-homophily network, where units with similar \(X_{i,1}\) are more likely to be connected. Network (ii) is a two-community network defined by \(X_{i,1}>0\) and \(X_{i,1}\leq 0\). Network (iii) is a latent-homophily network driven by both \(X_{i,1}\) and \(U_i\).
 
The outcome \(Y_i\) is generated by

(a): \(Y_i=1+X_{i,1}+X_{i,2}+2\bar{A}_{\mathcal{N}_i}+A_i\{1+X_{i,1}+2X_{i,2}-2\bar{A}_{\mathcal{N}_i}(1+X_{i,1}-X_{i,2})\}+\varepsilon_i\), where \(\varepsilon_i\) follows a standard normal distribution;

(b): \(Y_i=1+X_{i,1}^2+X_{i,2}+A_i(1+2X_{i,1}+X_{i,2}-3\bar{A}_{\mathcal{N}_i})+\varepsilon_i\).

We study all combinations of network generation (i)--(iii) and outcome generation (a)--(b). Using a Monte Carlo method, we find that the Nash regime exists in all of these settings. 

We compare the following methods.
(1) Our proposed method, denoted as Nash learning. We use tensor-product cubic B-splines as the sieve basis \(\Phi(X)\). For the unknown covariate-dependent propensity score setting, we estimate the propensity score using a correctly specified stratified model. 
(2) WOLS \cite[]{jiang2023dynamic} with neighbors' treatments assumed to be 1, denoted as WOLS1.
(3) WOLS with neighbors' treatments assumed to be 0, denoted as WOLS0.
(4) A value-search method that maximizes the overall counterfactual mean, inspired by \cite{viviano2025policy} and denoted as O-learning.
More details about the implementation of these methods can be found in Section \ref{Sec:competing} of the supplementary material.

The simulations are conducted with a sample size of $n=200$, with each setting replicated 200 times. In this study, we construct a test sample of $m=10,000$ units to assess the performance of the proposed methods. The evaluation metrics include the misclassification rate (MR), the regret, and the value. For a regime \(d(\cdot)\), the MR is defined on the test sample as
$
\frac{1}{m}\sum_{i=1}^m\left|{d}(X_i)-A_i^*\right|,
$
where $A_i^*$ denotes the Nash equilibrium treatment obtained by applying Algorithm \ref{alg:2} on the test samples with \(\hat{\tau}\) and \(\widehat{\mathcal{R}}\) replaced by their Monte Carlo counterparts. The regret is
$-\frac{1}{m}\sum_{i=1}^m[\min\{d(X_i){\tau}(X_i;d),0\}+\min\{(d(X_i)-1){\tau}(X_i;d),0\}]
$ with \(\tau\) calculated by Monte Carlo method. The value is $
\frac{1}{m}\sum_{i=1}^m \{\mu(X_i,\bar{d}_{\mathcal{N}_i})+d(X_i)\delta(X_i,\bar{d}_{\mathcal{N}_i})\}$. The MR measures disagreement with the reference Nash regime obtained by Monte Carlo computation, whereas the regret directly measures deviation from individual optimality. Smaller MR and regret indicate better performance. The value, in contrast, measures population-level performance.

We present the results in Tables \ref{tab1} and \ref{tab2} for covariate-dependent and constant propensity scores, respectively. Nash learning achieves the smallest MR and regret across all considered settings, consistent with its objective of targeting individual optimality. In contrast, O-learning generally achieves the highest population value, as it directly optimizes the overall counterfactual mean. Importantly, although Nash learning does not directly optimize population value, it achieves competitive population-level performance.

\begin{table}[!htbp]
\setlength{\tabcolsep}{3pt}
\caption{Mean of Monte Carlo MR, regret, and value (with standard deviation in parentheses) under a covariate-dependent propensity score. {The best result in each scenario is highlighted in bold.}}
\centering
\resizebox{\columnwidth}{!}{
\begin{tabular}{c c ccc ccc}
\toprule
& & \multicolumn{3}{c}{Outcome (a)}
& \multicolumn{3}{c}{Outcome (b)} \\
\cmidrule(lr){3-5}\cmidrule(lr){6-8}
Network & Method
& MR & Regret & Value
& MR & Regret & Value \\
\midrule

\multirow{4}{*}{(i)}
& Nash learning&$\boldsymbol{0.039}$ ($0.016$) & $\boldsymbol{0.021}$ ($0.020$) & $3.149$ ($0.051$)&$\boldsymbol{0.084}$ ($0.053$) & $\boldsymbol{0.060}$ ($0.097$) & $2.520$ ($0.092$)\\
& WOLS1&$0.120$ ($0.030$) & $0.140$ ($0.062$) & $2.908$ ($0.123$)&$0.257$ ($0.079$) & $0.548$ ($0.301$) & $2.515$ ($0.140$)\\
& WOLS0&$0.211$ ($0.054$) & $0.488$ ($0.239$) & $2.961$ ($0.156$)&$0.281$ ($0.112$) & $0.617$ ($0.425$) & $1.742$ ($0.493$)\\
& O-learning&$0.150$ ($0.049$) & $0.172$ ($0.104$) & $\boldsymbol{3.296}$ ($0.032$)&$0.165$ ($0.098$) & $0.272$ ($0.325$) & $\boldsymbol{2.578}$ ($0.147$)\\
\midrule

\multirow{4}{*}{(ii)}
& Nash learning&$\boldsymbol{0.035}$ ($0.017$) & $\boldsymbol{0.013}$ ($0.012$) & $3.157$ ($0.058$)&$\boldsymbol{0.108}$ ($0.063$) & $\boldsymbol{0.091}$ ($0.160$) & $2.488$ ($0.137$)\\
& WOLS1&$0.121$ ($0.031$) & $0.143$ ($0.059$) & $2.933$ ($0.129$)&$0.258$ ($0.062$) & $0.500$ ($0.256$) & $\boldsymbol{2.516}$ ($0.136$)\\
& WOLS0&$0.197$ ($0.064$) & $0.472$ ($0.272$) & $2.925$ ($0.199$)&$0.280$ ($0.081$) & $0.544$ ($0.273$) & $1.769$ ($0.298$)\\
& O-learning&$0.162$ ($0.048$) & $0.194$ ($0.112$) & $\boldsymbol{3.312}$ ($0.031$)&$0.141$ ($0.053$) & $0.158$ ($0.161$) & $2.476$ ($0.136$)\\
\midrule
\multirow{4}{*}{(iii)}
& Nash learning&$\boldsymbol{0.040}$ ($0.017$) & $\boldsymbol{0.025}$ ($0.023$) & $3.151$ ($0.049$)&$\boldsymbol{0.082}$ ($0.043$) & $\boldsymbol{0.055}$ ($0.060$) & $2.524$ ($0.102$)\\
& WOLS1&$0.117$ ($0.027$) & $0.135$ ($0.054$) & $2.928$ ($0.110$)&$0.264$ ($0.090$) & $0.557$ ($0.335$) & $2.513$ ($0.171$)\\
& WOLS0&$0.212$ ($0.053$) & $0.495$ ($0.229$) & $2.954$ ($0.150$)&$0.257$ ($0.087$) & $0.545$ ($0.340$) & $1.805$ ($0.418$)\\
& O-learning&$0.160$ ($0.054$) & $0.199$ ($0.125$) & $\boldsymbol{3.290}$ ($0.038$)&$0.172$ ($0.092$) & $0.269$ ($0.286$) & $\boldsymbol{2.597}$ ($0.124$)\\
\bottomrule
\end{tabular}
}
\label{tab1}
\end{table}

\begin{table}[!htbp]
\setlength{\tabcolsep}{3pt}
\caption{Mean of Monte Carlo MR, regret, and value (with standard deviation in parentheses) under a constant propensity score. {The best result in each scenario is highlighted in bold.}}
\centering
\resizebox{\columnwidth}{!}{
\begin{tabular}{c c ccc ccc}
\toprule
& & \multicolumn{3}{c}{Outcome (a)}
& \multicolumn{3}{c}{Outcome (b)} \\
\cmidrule(lr){3-5}\cmidrule(lr){6-8}
Network & Method
& MR & Regret & Value
& MR & Regret & Value \\
\midrule

\multirow{4}{*}{(i)}
& Nash learning&$\boldsymbol{0.032}$ ($0.015$) & $\boldsymbol{0.018}$ ($0.017$) & $3.156$ ($0.040$)&$\boldsymbol{0.075}$ ($0.035$) & $\boldsymbol{0.046}$ ($0.047$) & $2.528$ ($0.092$)\\
& WOLS1&$0.115$ ($0.025$) & $0.134$ ($0.056$) & $2.921$ ($0.095$)&$0.256$ ($0.089$) & $0.553$ ($0.354$) & $2.505$ ($0.169$)\\
& WOLS0&$0.211$ ($0.043$) & $0.480$ ($0.194$) & $2.971$ ($0.129$)&$0.250$ ($0.096$) & $0.503$ ($0.376$) & $1.849$ ($0.434$)\\
& O-learning&$0.153$ ($0.041$) & $0.170$ ($0.090$) & $\boldsymbol{3.304}$ ($0.027$)&$0.180$ ($0.079$) & $0.289$ ($0.239$) & $\boldsymbol{2.602}$ ($0.108$)\\
\midrule

\multirow{4}{*}{(ii)}
& Nash learning &$\boldsymbol{0.026}$ ($0.012$) & $\boldsymbol{0.008}$ ($0.007$) & $3.163$ ($0.039$)&$\boldsymbol{0.104}$ ($0.049$) & $\boldsymbol{0.078}$ ($0.085$) & $2.501$ ($0.098$)\\
& WOLS1&$0.117$ ($0.030$) & $0.142$ ($0.069$) & $2.933$ ($0.114$)&$0.263$ ($0.096$) & $0.584$ ($0.418$) & $2.448$ ($0.184$)\\
& WOLS0&$0.212$ ($0.051$) & $0.513$ ($0.232$) & $2.904$ ($0.186$)&$0.289$ ($0.096$) & $0.610$ ($0.402$) & $1.708$ ($0.435$)\\
& O-learning&$0.161$ ($0.045$) & $0.189$ ($0.105$) & $\boldsymbol{3.314}$ ($0.032$)&$0.187$ ($0.078$) & $0.287$ ($0.257$) & $\boldsymbol{2.537}$ ($0.133$)\\
\midrule
\multirow{4}{*}{(iii)}
& Nash learning&$\boldsymbol{0.035}$ ($0.015$) & $\boldsymbol{0.022}$ ($0.020$) & $3.149$ ($0.041$)&$\boldsymbol{0.069}$ ($0.038$) & $\boldsymbol{0.039}$ ($0.046$) & $2.539$ ($0.079$)\\
& WOLS1&$0.113$ ($0.024$) & $0.132$ ($0.051$) & $2.925$ ($0.087$)&$0.261$ ($0.089$) & $0.556$ ($0.376$) & $2.511$ ($0.181$)\\
& WOLS0&$0.208$ ($0.037$) & $0.471$ ($0.159$) & $2.970$ ($0.110$)&$0.252$ ($0.100$) & $0.553$ ($0.431$) & $1.778$ ($0.482$)\\
& O-learning&$0.151$ ($0.041$) & $0.169$ ($0.094$) & $\boldsymbol{3.302}$ ($0.032$)&$0.178$ ($0.075$) & $0.260$ ($0.215$) & $\boldsymbol{2.617}$ ($0.106$)\\
\bottomrule
\end{tabular}
}
\label{tab2}
\end{table}

In the supplementary material Section \ref{Sec:addsimulation}, we present additional simulation studies that consider a larger number of covariates and a range of sample sizes. We also report additional metrics including computational time and the estimated regret given the estimated OITR. Notably, our proposed Nash learning method demonstrates greater computational efficiency than O-learning. Moreover, we include additional outcome-generating mechanisms, including one simulation in which the existence of a Nash regime can be established using Theorem \ref{the:nashexistence}, as well as another under which the Nash regime does not exist in the supplementary material.
In addition, we evaluate the doubly robust property of our method under misspecification of either the propensity score \(\pi\) or the baseline outcome function \(\mu\).



\subsection{Results across different network sparsity levels} \label{Sec:diffsparsity}
In this subsection, we examine how the empirical performance of Nash learning varies with network sparsity and compare the observed patterns with Theorems \ref{the:asymptotic1} and \ref{the:asymptotic2}.

\begin{figure}[!htbp]
  \centering      \includegraphics[width=\textwidth]{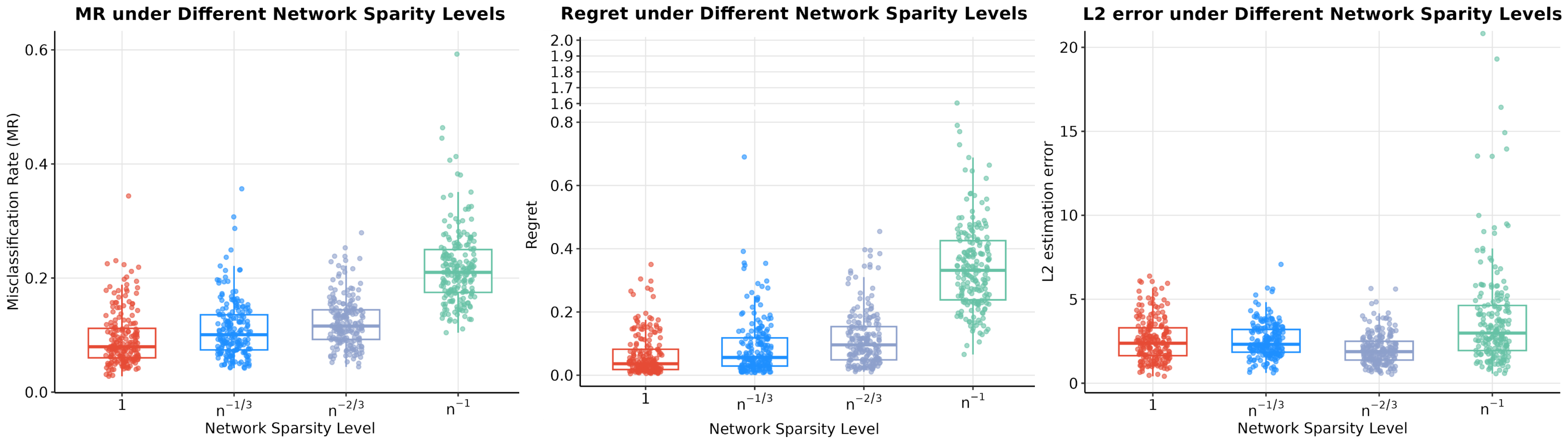}
 \caption{Boxplots of Monte Carlo MR, Monte Carlo regret, and the \(L_2\) estimation error \(\|\hat{\beta}-\beta^*\|_2\) of our method under a covariate-dependent propensity score and different network sparsity levels.}\label{figure:sparsitynonconstanttheory}
\end{figure}

We first investigate a covariate-dependent propensity score. The treatment \(A_i\) is generated by the propensity score \(\pi(A_i=1|X_i)=0.8I(X_{i,1}>0)+0.2I(X_{i,1}\leq 0)\). We choose network generation function \(G_n(X_i,U_i,X_j,U_j)=\rho_n G(X_i,U_i,X_j,U_j)\) with \(\rho_n=1, n^{-1/3},n^{-2/3}\) and \(n^{-1}\). Recall that a smaller \(\rho_n\) corresponds to a sparser network. The function \(G(X_i,U_i,X_j,U_j)\) is chosen as (ii) in Section \ref{Sec:simconstantps}. The outcome generation follows (b) of Section \ref{Sec:simconstantps}. Other data-generating processes follow Section \ref{Sec:simconstantps}. We provide boxplots of the MR, regret, and the \(L_2\) estimation error \(\|\hat{\beta}-\beta^*\|_2\) for our Nash learning in Figure \ref{figure:sparsitynonconstanttheory}.

  

Figure \ref{figure:sparsitynonconstanttheory} shows that the boxplots for \(\rho_n=n^{-1}\) are substantially higher than those for the other sparsity levels. Moreover, MR and regret increase as the network becomes sparser. These patterns are consistent with Theorem \ref{the:asymptotic1}, since consistency is not guaranteed when \(\rho_n=n^{-1}\), and the convergence of regime estimation depends on the effective neighborhood size \(n\rho_n\). The \(L_2\) estimation error of \(\beta^*\) decreases slightly as the network becomes sparser when \(\rho_n\neq n^{-1}\). This is also reasonable because, although Theorem \ref{the:asymptotic1}(i) establishes root-\(n\) consistency of \(\hat\beta\) without dependence on network sparsity \(\rho_n\), its finite-sample performance can still be affected by \(\rho_n\).

\begin{figure}[!htbp]
  \centering      
  \includegraphics[width=\textwidth]{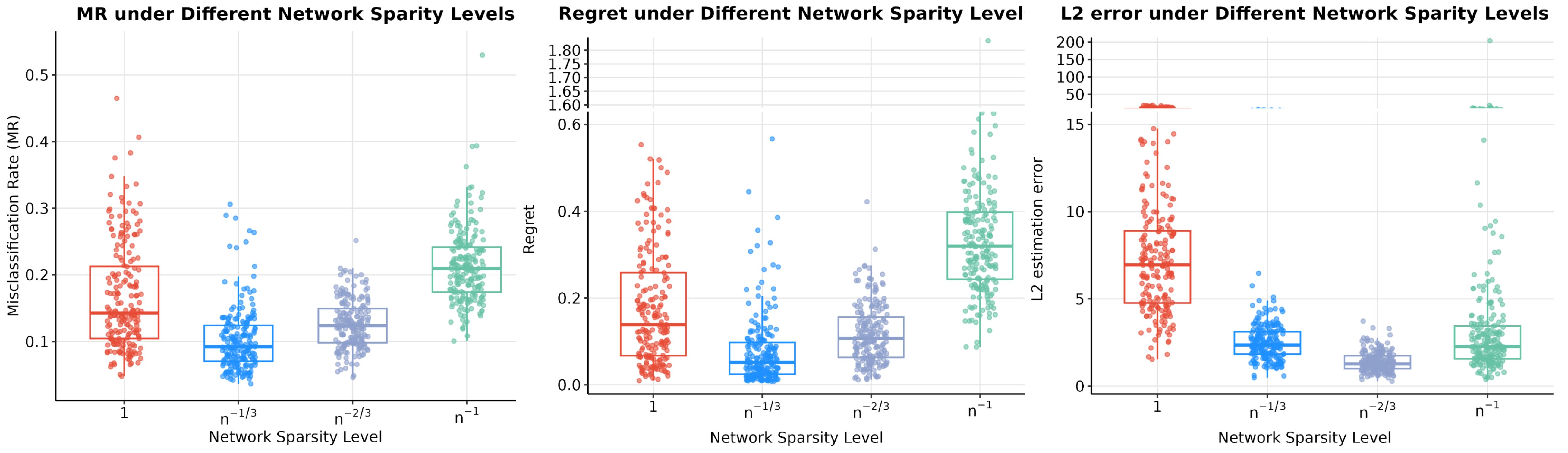}
 \caption{Boxplots of Monte Carlo MR, Monte Carlo regret, and the \(L_2\) estimation error \(\|\hat{\beta}-\beta^*\|_2\) of our method under a constant propensity score and different network sparsity levels. 
}\label{figure:sparsityconstanttheory}
\end{figure}

We then investigate a constant propensity score. We use the same data-generating mechanism as above, except that treatment is assigned with the constant propensity score \(\pi(A_i=1\mid X_i)=0.5\). We provide boxplots of the MR, regret, and the \(L_2\) estimation error \(\|\hat{\beta}-\beta^*\|_2\) for our Nash learning in Figure \ref{figure:sparsityconstanttheory}.

Figure \ref{figure:sparsityconstanttheory} displays a clear U-shaped pattern across different sparsity levels, with the boxplots for \(\rho_n=1\) and \(\rho_n=n^{-1}\) positioned substantially higher than those at intermediate sparsity levels. This pattern is consistent with the near-collinearity mechanism underlying Theorem \ref{the:asymptotic2}.

\section{Real-data application} \label{Sec:realdata}
In this section, we illustrate an application of our Nash learning using a network dataset from \cite{cai2015social} to study how to assign information sessions to farmers in order to improve their knowledge of disaster insurance. Assigning information sessions in a way that achieves individual optimality is important, because farmers who are assigned a session type that is not individually optimal given their peers' assignments may have incentives to deviate from the prescribed assignment. The resulting Nash regime is therefore more stable and policy-relevant. 

The dataset comprises 4,569 households distributed across 47 distinct villages in a rural region of Jiangxi Province, China. The social network \(\boldsymbol{E}\) is constructed from survey-based friendship nominations. Two households are considered connected if either household nominates the other as a friend. 
Since household connections in different villages are exceedingly rare, we exclude them, thereby yielding 47 distinct clusters within the social network. The cluster structure facilitates the evaluation of our method via repeated random cluster splitting.

The original study evaluated two types of information sessions: a simple session introducing disaster insurance and an intensive session providing more advanced information. The outcomes of interest in the original study included farmers' insurance knowledge and insurance take-up. Insurance knowledge was measured by a test score ranging from 0 to 10 based on questions about disaster insurance. In our analysis, we use insurance knowledge as the outcome \(Y_i\). 

In addition to the random assignment of information sessions, households in the study were also randomly allocated to either round 1 or round 2 with equal probability, with round 2 sessions scheduled a few days after those in round 1. This temporal gap facilitates more extensive knowledge diffusion for farmers in round 1 through their social networks. Consequently, four groups emerge: round 1 with an intensive session; round 2 with an intensive session; round 1 with a simple session; and round 2 with a simple session. Various combinations of these groups can be examined. Following \cite{Bargagli2020Heterogeneous}, we adopt a one-versus-rest strategy to investigate the assignment of the round 1 intensive session ($A_i = 1$) relative to all others ($A_i = 0$).

We include observed characteristics that may plausibly capture treatment effect heterogeneity as covariates \(X_i\) for constructing the optimal regime. These include binary indicators of whether the household head has an education level higher than primary school and whether the household has ever received an insurance payout, as well as several continuous or count variables, including a measure of risk aversion and household size.

We employ repeated random cluster splitting to evaluate competing methods. Specifically, we randomly partition the 47 clusters into training and test sets, allocating 50\% of the clusters to the former and the remainder to the latter. 
Within the training sets, we implement each of the competing methods described in Section \ref{Sec:simulation} to estimate the optimal regime. For our Nash learning method, we use tensor-product bases constructed from indicator basis functions for binary covariates and cubic B-splines for continuous or count covariates.
The estimated regime is then evaluated on both the training and test sets using estimated regret and estimated value as performance metrics. These metrics are defined analogously to the regret and value metrics in Section \ref{Sec:simconstantps}, except that the Monte Carlo quantities \(\mu\), \(\delta\), and \(\tau\) are replaced by their estimates obtained from the training data for training-set evaluation and from the test data for test-set evaluation.


The cluster splitting procedure is repeated 200 times. We report the averages of the estimated regret and estimated value over both the training and test datasets in Table \ref{real_data}. Our method achieves a very small estimated regret on the training set, indicating that the fitted regime has very small empirical deviation from individual optimality. In addition, O-learning performs well in terms of the estimated value on the training set as expected. For the metrics evaluated on the test set, our Nash learning method outperforms the competing approaches in both estimated regret and estimated value. The estimated regret on the test set is not as close to zero as that on the training set, which is likely attributable to minor distributional shifts across clusters. 

 \begin{table}[!htbp]
    \caption{Mean (standard deviation) of estimated value and estimated regret on both the training set and test set for competing methods. {The best result in each scenario is highlighted in bold.} } \label{real_data}
    \centering
    \resizebox{0.8\columnwidth}{!}{
  \begin{tabular}{lcccc}
  \toprule
  & \multicolumn{2}{c}{Metrics on the training set} & \multicolumn{2}{c}{Metrics on the test set}\\
 \cmidrule(lr){2-3}\cmidrule(lr){4-5}
 Method  & Estimated regret & Estimated value & Estimated regret & Estimated value \\
  \midrule Nash learning & \(\boldsymbol{0.001}\) \((0.001)\)  & \(5.710\) \((0.354)\) & \(\boldsymbol{0.171}\) \((0.092)\) & \(\boldsymbol{5.497}\) \((0.312)\) \\
  WOLS1 & \({0.009}\) \((0.015)\)  & \({5.702}\) \((0.352)\) & \(0.199\) \((0.122)\) & \(5.459\) \((0.286)\) \\
  WOLS0 & \(0.267\) \((0.157)\)  & \(5.427\) \((0.380)\) & \(0.239\) \((0.140)\) &\(5.458\) \((0.407)\) \\
  O-learning & \(0.017\) \((0.025)\)  & \(\boldsymbol{5.713}\) \((0.357)\) & \(0.191\) \((0.107)\) & \(5.472\) \((0.309)\) \\
  \bottomrule
  \end{tabular}
  }
\end{table}

\section{Discussion}\label{Sec:discussion}
In this paper, we introduce Nash regimes into the OITR literature as a new decision target that emphasizes individual optimality. Under a Nash regime, no individual can improve their conditional expected outcome through unilateral deviation. Because an exact Nash regime may not always exist, we further introduce an \(\alpha^*\)-Nash regime, which minimizes the attainable regret relative to individual optimality. We then propose Nash learning to estimate the proposed regimes. Simulation studies and a real-data analysis show that Nash learning achieves low individual-level regret while maintaining competitive population value.



\section*{Acknowledgments}
{ChatGPT was used for English polishing and coding assistance. The authors reviewed all AI-assisted content and take full responsibility.}

\section*{Conflict of interest}

The authors declare that they have no conflicts of interest.

\section*{Funding}

Junwen Xia and Jingxiao Zhang were partially supported by the MOE Project of Key Research Institute of Humanities and Social Sciences (22JJD910001).
Dehan Kong was partially supported by NSERC Discovery Grants (RGPIN-2022-04646).

\section*{Data availability}

The network dataset that supports the findings of this study is publicly available and can be downloaded from \url{https://www.openicpsr.org/openicpsr/project/113593/version/V1/view}.

\bibliography{reference}

\end{document}